\documentclass[a4paper]{article}

\begin {document}

\title {\bf Band theory in the context of the 
Hamilton-Jacobi formulation }
\author{A.~Bouda\footnote{Electronic address: 
{\tt bouda\_a@yahoo.fr}} 
\ and A.~ Mohamed Meziane\footnote{Electronic address: 
{\tt amohamed\_meziane@yahoo.fr}}\\
Laboratoire de Physique Th\'eorique, Universit\'e de B\'eja\"\i a,\\ 
Route Targa Ouazemour, 06000 B\'eja\"\i a, Algeria\\}

\date{\today}

\maketitle

\begin{abstract}
\noindent
In the one-dimensional periodic potential case, we formulate the condition 
of Bloch periodicity for the reduced action by using the relation between 
the wave function and the reduced action established in the context of 
the equivalence postulate of quantum mechanics. Then, without 
appealing to the wave function properties, we reproduce the 
well-known dispersion relations which predict the band structure 
for the energy spectrum in the Kr\"onig-Penney model.
\end{abstract}

\vskip\baselineskip



\newpage

\section{Introduction}

By establishing that quantum mechanics can be reproduced 
from an equivalence postulate \cite{FM1a,FM1b,FM2}, Faraggi and Matone
have rekindled the hope that general relativity can be reconciled 
with quantum mechanics. Assuming that all quantum systems can be connected 
by a coordinate transformation, they derived the one-dimensional 
quantum stationary Hamilton-Jacobi equation
\begin {eqnarray}
{1\over 2m} \left({\partial S_0 \over \partial x}\right)^2 + V(x)-E= 
\hskip60mm&& \nonumber\\
{\hbar^2\over 4m}  \left[{3\over 2}\left( 
{\partial S_0 \over\partial x}\right)
^{- 2 }\left({\partial^2 S_0 \over \partial x^2}\right)^2-
\left( {\partial S_0 \over \partial x  }\right)^{- 1 }
{\partial^3 S_0 \over \partial x^3  } \right] \; ,
\end {eqnarray}
in which $V(x)$ is an external potential and $E$ the energy. 
They established then that the Schr\"odinger wave function is 
related to the reduced action, $S_0$, by 
\begin {equation}
\phi(x)= R(x)
\left[\alpha\ \exp\left({i\over\hbar }S_0(x)\right)
+\beta\ \exp\left(-{i\over\hbar }S_0(x)\right)\right]\; ,
\end {equation} 
as shown also in \cite{B1} by using the probability current. 
In Eq. (2), $\alpha$ and $\beta$ are 
complex constants, $S_0(x)$ and $R(x)$ are real functions and 
$R(x)$ is proportional to $({\partial S_0 / \partial x})^{-1/2}$. 
In contrast to Bohm's theory \cite{Bohm1,Bohm2}, relation (2) guarantees 
that $S_0$ is never constant even in the case where the wave 
function is real, up to a constant phase factor. We note that 
the Bohm ansatz is obtained from (2) by using the particular values 
$\alpha=1$ and $\beta=0$. 

Furthermore, without appealing to the usual axiomatic 
interpretation of the wave function, Faraggi and Matone
\cite{FM2, FM3} showed that tunnel effect and energy quantization 
are consequences of the equivalence postulate. 
In the same spirit, we propose in this paper to examine the case of 
a system in a periodic potential. In section 2, we establish  
the condition of Bloch periodicity \cite{Bloch} for the reduced 
action. In section 3, we investigate the Kr\"onig-Penney model \cite{KP} 
without appealing to the Schr\"odinger wave function or to its usual
axiomatic interpretation. Section 4 is devoted to conclusion.

\section{The Bloch theorem}

 The understanding of the behavior of electrons in 
crystal lattices has been advanced through the work of Bloch 
\cite{Bloch}. The main idea is that the interaction of an  
electron with the other particles of the lattice may be replaced 
by a periodic potential.  

In the present work, we consider the one-dimensional case with 
a potential satisfying the following periodicity condition
\begin {equation}
V(x+e)=V(x) , \ \ \ \ \ \ \   \forall \ x \ ,
\end {equation} 
where $e$ is a period. With this relation, Bloch \cite{Bloch} 
showed that any solution $\phi$ of the Schr\"odinger equation,
\begin {equation}
-{\hbar^2 \over 2m} {d^2 \phi \over dx^2 } + V(x) \phi = E \phi \ ,
\end {equation}
satisfies the property
\begin {equation}
\phi (x+e) =  \exp{(iKe)} \, \phi(x) \ ,
\end {equation}
where $K$ is a constant. This property represents the condition of 
Bloch periodicity for the wave function. It is  known as a 
Bloch theorem and was also established by Floquet \cite{Floquet}.

Our task now consists in finding a corresponding version 
when we deal with the reduced action which is related to the 
wave function by (2). For this purpose, let us set 
\begin {equation}
\alpha = |\alpha|\ \exp{(ia)} \ , \hskip20mm 
\beta = |\beta| \ \exp{(ib)}\ ,
\end {equation}
$a$ and $b$ being real parameters. 
By substituting expressions (6) in (2), we can deduce that
\begin {eqnarray}
\exp{(iKe)} \ \phi(x)= \exp{\left( i{a+b \over 2}\right)} R(x) 
              \hskip55mm&& \nonumber\\
         \left\{ \left[(|\alpha| +|\beta|) \ \cos \left({S_0(x)\over\hbar }
             +{ a-b\over 2}\right) \cos Ke \right. \right.
         \hskip25mm&& \nonumber\\
           \left. -(|\alpha| -|\beta|) \ 
                 \sin \left({S_0(x)\over\hbar}+{ a-b\over 2} \right) 
                  \sin Ke \right]
             \hskip12mm&& \nonumber\\
           + i \left[ (|\alpha| -|\beta|) \ \sin \left({S_0(x)\over\hbar }
             +{ a-b\over 2}\right) \cos Ke \right.
         \hskip20mm&& \nonumber\\
          \left. \left. + (|\alpha| +|\beta|)   
             \cos \left({S_0(x)\over\hbar }+{ a-b\over 2} \right)
              \sin Ke \right] \right\}
\end {eqnarray} 
Writing an analogous relation for $\phi(x+e)$ as in (2) and using (6), 
we obtain
\begin {eqnarray}
\phi(x+e)= \exp{\left( i{a+b \over 2}\right)}  R(x+e)
            \hskip55mm&& \nonumber\\
          \left[(|\alpha| +|\beta|) \ \cos \left({S_0(x+e)\over\hbar }
             +{ a-b \over 2}\right) \right. \hskip25mm&& \nonumber\\
        + \left. i(|\alpha| -|\beta|) \ \sin \left({S_0(x+e)\over\hbar }
             +{ a-b\over 2}\right) \right]
\end {eqnarray} 
Substituting (7) and (8) in (5) and separating in the obtained relation 
the real part from the imaginary one, we get to the 
two following relations
\begin{eqnarray}
\left(|\alpha|+|\beta|\right) R\left(x+e\right)
    \cos\left({S_{0}\left(x+e\right)\over\hbar}+{a-b\over2}\right)
     =\nonumber\hskip 3.75cm\\
      R\left(x\right)\left[\left(|\alpha|+|\beta|\right)
      \cos\left({S_{0}\left(x\right)\over\hbar}+{a-b\over2}\right)
      \cos Ke \right.\hskip 2cm\\
      \left.-\left(|\alpha|-|\beta|\right)
      \sin\left({S_{0}\left(x\right)\over\hbar}+{a-b\over2}\right)
      \sin Ke\right] \nonumber
\end{eqnarray}
and
\begin{eqnarray}
\left(|\alpha|-|\beta|\right) R\left(x+e\right)
        \sin\left({S_{0}\left(x+e\right)\over\hbar}+{a-b\over2}\right)
        =\hskip 3.75cm\nonumber\\
         R\left(x\right)\left[\left(|\alpha|-|\beta|\right)
         \sin\left({S_{0}\left(x\right)\over\hbar}+{a-b\over2}\right)
         \cos Ke \right.\hskip 1.5cm\\
         \left.+\left(|\alpha|+|\beta|\right)
         \cos\left({S_{0}\left(x\right)\over\hbar}+{a-b\over2}\right)
         \sin Ke \right].\nonumber
\end{eqnarray}
Dividing side by side relations (10) and (9), we obtain
\begin{equation}
\Gamma\tan\left[{S_{0}\left(x+e\right)\over\hbar}+\Delta\right]
          ={\Gamma\tan\left[{S_{0}\left(x\right) / \hbar}
            +\Delta\right]+\tan{Ke}\over 
            1-\Gamma\tan\left[{S_{0}\left(x\right) / \hbar}+
            \Delta\right]\tan{Ke}}
\end{equation}
where
\begin{equation}
\Delta={a-b\over2} \ , \hskip20mm \Gamma
       ={|\alpha|-|\beta|\over|\alpha|+|\beta|} \ .
\end{equation}
Knowing that $\tan \ (\arctan u) = u$ $\forall \  u \in \Re$, 
with the use of 
\[
u=\Gamma \tan \left[{{S_{0}(x)\over \hbar} +\Delta}\right] \ , 
\]
relation (11) turns out to be
\begin{eqnarray}
\arctan\left\{\Gamma\tan\left[{S_{0}\left(x+e\right)\over\hbar}
     +\Delta\right]\right\}= \hskip35mm \nonumber\\
     \arctan\left\{\Gamma\tan\left[{S_{0}\left(x\right)\over\hbar}
     +\Delta\right]\right\}+Ke+n\pi
\end{eqnarray}
where $n$ is an integer number. This relation is the condition  
of Bloch periodicity for the reduced action and represents 
the Bloch theorem version in this context. Taking into account 
the relation $\tan \alpha = -i [\exp(2i\alpha)-1][\exp(2i\alpha)+1]^{-1}$, 
it is easy to show from (11) that the periodicity condition (13) can be 
written in the following form
\begin{equation}
\exp[2iS_0(x+e)/\hbar]=\frac{P\exp[2iS_0(x)/\hbar]+Q}{M\exp[2iS_0(x)/\hbar]+N} \ ,
\end{equation}
where 
\begin{eqnarray}
P& = &-(1-\Gamma)^2+(1+\Gamma)^2 \exp(2iKe) \ , \\
Q& = &(1-\Gamma^2)[\exp(2iKe)-1]\exp(-2i\Delta) \ , \\
M& = &-(1-\Gamma^2)[\exp(2iKe)-1]\exp(2i\Delta) \ , \\
N& = &(1+\Gamma)^2-(1-\Gamma)^2 \exp(2iKe) \ .
\end{eqnarray}
Relation (14) indicates that $\exp[2iS_0(x+e)/\hbar]$ is the  
M\"obius transformation of $\exp[2iS_0(x)/\hbar]$. The M\"obius group 
has allowed to fix from the equivalence postulate the quantum version 
of the Hamilton-Jacobi equation \cite{FM2}. 
The trace of the M\"obius transformation (14) is $P+N=4\Gamma[1+\exp(2iKe)]$.
Except for the particular values of $K$ with which $\sin Ke $ vanishes, this 
trace is complex and hence the transformation (14) can not be classified 
as hyperbolic, parabolic or elliptic \cite{DNF}.
In the case of Bohm's theory, we have the particular values 
$\alpha=1$ and $\beta =0$ which imply that $\Gamma=1$ and $\Delta=0$. 
It follows that both relations (13) and (14) reduce to
\begin{equation}
S_{0}\left(x+e\right)=S_{0}\left(x\right)+\hbar Ke + n' \pi \hbar \ , 
\end{equation}
where $n'$ is also an integer number. It is interesting to observe that if 
we define the function
\begin{equation}
F(x)\equiv \frac{1}{\pi\hbar} [S_o(x) -\hbar Kx] \ ,
\end{equation}
we can show from (19) the following affine transformation 
\begin{equation}
F(x+e)=F(x)+n' \ .
\end{equation}
%
%

\section{The Kr\"onig-Penney model}

Another important step in the description of the behavior 
of electrons in crystal lattices was the work of Kr\"onig and Penney 
\cite{KP}. In one dimension, their model, which has the advantage in 
that it predicts correctly the spectrum of permissible 
energy values, consists in considering the potential in the form of a 
series of equidistant rectangular barriers
\[
V(x) = \left\{ \begin{array}{cc}
              0, & \  \  n(c+d)< x < n(c+d)+c \\ [.1in]
              V_0, & \  \  n(c+d)+c < x < (n+1)(c+d) 
               \end{array} 
       \right. ,
\label{eq:ve}
\]
where $n$ is an integer number. The period is $e=c+d$. 

Let us begin by the case where $E>V_{0}$ and set 
\begin{equation}
k_{1}={\sqrt{2m(E-V_{0})} \over \hbar} \ , 
       \hskip 15mm k_{2}={\sqrt{2mE} \over \hbar} \ .
\end{equation}
In Refs. \cite{NGO,Sin}, by using the continuity of the wave function 
and its derivative, it is shown that 
\begin{equation}
\cos Ke =
       \cos\left(k_{1}d\right)\cos\left(k_{2}c\right)-
        {{k_{1}^2+k_{2}^2}\over{2k_{1}k_{2}}}
        \sin\left(k_{1}d\right)\sin\left(k_{2}c\right)
\end{equation}
An investigation of this dispersion relation shows the 
existence of a band structure for the energy spectrum.

Our goal now is to reproduce relation (23) by using the 
properties of the reduced action. 

Let us call $I$, $II$ and $III$ the three regions $-d<x<0$, $0<x<c$ 
and $c<x<c+d$ respectively and impose the continuity at $x=0$ for 
the reduced action and its first and second derivative
\begin{equation}
\left. S_0^I(x)\right\vert_0=\left. S_0^{II}(x)\right\vert_0 \ ,
\end{equation}
\begin{equation}
\left. {\partial S_0^I(x)\over \partial x} \right\vert_0
=\left. {\partial S_0^{II}(x)\over \partial x} \right\vert_0 \ ,
\end{equation}
\begin{equation}
\left. {\partial^2 S_0^I(x)\over \partial x^2} \right\vert_0
=\left.{\partial^2 S_0^{II}(x)\over \partial x^2} \right\vert_0 \ .
\end{equation} 
Since $c=e-d$, by assuming at $x=-d$ the following continuity  
condition: 
\[
\left. S_{0}^{III}(x+e)\right\vert_{x=-d} = 
\left. S_{0}^{II}(x+e)\right\vert_{x=-d}, 
\]
and by applying at $x=-d$ the condition of Bloch periodicity, 
Eq. (13), for the reduced action, we deduce that 
\begin{eqnarray}
\left. \arctan\left\{\Gamma\tan\left[{S_{0}^{II}\left(x+e\right)
     \over\hbar}
     +\Delta\right]\right\}
     \right\vert_{x=-d}= \hskip35mm \nonumber\\
     \left. \arctan\left\{
       \Gamma\tan\left[{S_{0}^{I}\left(x\right)\over\hbar} +\Delta\right]
       \right\}
      \right\vert_{x=-d} +Ke+n\pi \ .
\end{eqnarray}
As at $x=0$, by assuming also the continuity at $x=-d$ of the 
first and the second derivative of $S_{0}(x+e)$, 
\[
\left. {\partial S_0^{III}(x+e)\over \partial x} \right\vert_{x=-d}
=\left. {\partial S_0^{II}(x+e)\over \partial x} \right\vert_{x=-d} \ ,
\]
\[
\left. {\partial^2 S_0^{III}(x+e)\over \partial x^2} \right\vert_{x=-d}
=\left.{\partial^2 S_0^{II}(x+e)\over \partial x^2} \right\vert_{x=-d} \ ,
\]
we can take the first and the second 
derivative of the two sides of relation (27) 
\begin{eqnarray}
\left. {\partial \over \partial x} 
\arctan\left\{\Gamma\tan\left[{S_{0}^{II}\left(x+e\right)\over\hbar}
     +\Delta\right]\right\}\right\vert_{x=-d}= \hskip35mm \nonumber\\
     \left. {\partial \over \partial x}
     \arctan\left\{\Gamma\tan\left[{S_{0}^{I}\left(x\right)\over\hbar}
     +\Delta\right]\right\}\right\vert_{x=-d} \ ,
\end{eqnarray}
\begin{eqnarray}
\left.{\partial^2 \over \partial x^2} 
\arctan\left\{\Gamma\tan\left[{S_{0}^{II}\left(x+e\right)\over\hbar}
     +\Delta\right]\right\}\right\vert_{x=-d}= \hskip35mm \nonumber\\
      \left. {\partial^2 \over \partial x^2}
     \arctan\left\{\Gamma\tan\left[{S_{0}^{I}\left(x\right)\over\hbar}
     +\Delta\right]\right\}\right\vert_{x=-d} \ .
\end{eqnarray}
The solution for the one-dimensional quantum stationary 
Hamilton-Jacobi equation, Eq. (1), is well-known 
\cite{FM1a,FM2,B1,Fl1,Fl2,Fl3} and is written in \cite{BD1} as  
\begin {equation}
S_0=\hbar \ \arctan {\left [ \mu  {\phi_1 
\over \phi_2 } +\nu \right ]} +\hbar l \; ,
\end {equation}
where $(\phi_1,\phi_2)$ is a couple of two real independent solutions
of the Schr\"odinger equation, Eq. (4), and $(\mu,\nu,l)$ are 
real integration constants satisfying the condition $\mu \ne 0$. 
Let us choose for Eq. (4) the couples of independent solutions
\begin{equation}
(\sin k_1x, \cos k_1x), \hskip20mm (\sin k_2x, \cos k_2x)
\end{equation}
respectively in regions $I$ and $II$. With the use of (30), we have
\begin {equation}
S_0^I(x)=\hbar \ \arctan {\left[ \mu_1 \tan (k_1x)  +\nu_1  \right]} 
           +\hbar l_1 \; ,
\end {equation}
and
\begin {equation}
S_0^{II}(x)=\hbar \ \arctan {\left[ \mu_2 \tan (k_2x)  +\nu_2  \right]} 
         +\hbar l_2 \; .
\end {equation}
As the reduced action is defined up to an additive constant, we can 
fix one constant among $(l_1,l_2)$ and determine the other from 
the boundary conditions. Thus, let us  choose
\begin {equation}
 l_1 = -\Delta\; ,
\end {equation}
where $\Delta$ is defined in (12), and apply relations (24), (25) 
and (26)
\begin {equation}
\hbar \ \arctan {(\nu_1)}-\hbar \Delta 
      = \hbar \ \arctan {( \nu_2  )}+\hbar l_2 \; ,
\end {equation}
\begin {equation}
\hbar {\mu_1 k_1 \over {1+\nu_1^2}}
      = \hbar {\mu_2 k_2 \over {1+\nu_2^2}} \; ,
\end {equation}
\begin {equation}
- \hbar {2 \mu_1^2 \nu_1 k_1^2 \over {(1+\nu_1^2)^2}}
      = - \hbar {2 \mu_2^2 \nu_2 k_2^2 \over {(1+\nu_2^2)^2}} \; .
\end {equation}
From the system (35), (36) and (37), it is easy to show that
\begin {eqnarray}
\nu_1 & = & \nu_2 \ ,  \\
 l_1& = & l_2= -\Delta  \ ,\\
 \mu_1  & = &  {k_2 \over k_1} \mu_2 \; ,
\end {eqnarray}
and relations (32) and (33) become
\begin {equation}
S_0^I(x)=\hbar \ \arctan {\left[ \mu_1 \tan (k_1x)  +\nu_1  \right]} 
           -\hbar \Delta \; ,
\end {equation}
and
\begin {equation}
S_0^{II}(x)=\hbar \ \arctan {\left[ {k_1 \over k_2 }\mu_1 \tan (k_2x)  
+\nu_1 \right]} -\hbar \Delta \; .
\end {equation}
As $c=e-d$, if we set
\begin{eqnarray}
A & = & -\mu_{1}\tan\left(k_{1}d\right)+\nu_{1} \\
B & = & {{k_{1}}\over{k_{2}}}\mu_{1}\tan\left(k_{2}c\right)+\nu_{1}
\end{eqnarray}
relation (27) gives
\begin{equation}
\arctan\left\{\Gamma \tan\left[\arctan B \right]\right\}= 
\arctan\left\{\Gamma \tan\left[\arctan A \right]\right\}
+Ke+n\pi
\end{equation}
which leads to
\begin{equation}
\Gamma B = {\Gamma A +\tan Ke \over 1-\Gamma A \tan Ke } \ .
\end{equation}
This relation can be rewritten in the following form
\begin{equation}
\cos^2 Ke = {(1+ \Gamma^2 AB)^2 \over (1+ \Gamma^2 A^2)(1+ \Gamma^2 B^2)} \ .
\end{equation}
With the use of (41), (42), (43) and (44), by applying 
successively (28) and (29), we find
\begin{equation}
(1+ \Gamma^2 B^2) \cos^2 k_2c = (1+ \Gamma^2 A^2) \cos^2 k_1d 
\end{equation}
and
\begin{eqnarray}
{ k_2 \tan k_2c \over (1+ \Gamma^2 B^2) \cos^2 k_2c}
+ { k_1 \tan k_1d \over (1+ \Gamma^2 A^2) \cos^2 k_1d} 
            \hskip35mm\nonumber\\
=  { \mu_1 k_1 \Gamma^2 B \over (1+ \Gamma^2 B^2)^2 \cos^4 k_2c}
   - { \mu_1 k_1 \Gamma^2 A \over (1+ \Gamma^2 A^2)^2 \cos^4 k_1d}     \ .
\end{eqnarray}
Taking into account relations (48), (47) and (49) give respectively
\begin{equation}
\cos^2 Ke=
        \left({{1+\Gamma^{2}AB}
          \over{1+\Gamma^{2}A^2}}\right)^2
          {{\cos^2 k_{2}c}\over{\cos^2 k_1d}}
\end{equation}
and
\begin{equation}
 \Gamma^2 (B-A) = {1+ \Gamma^2 A^2 \over \mu_1 k_1  } 
       (k_2 \tan k_2c  + k_1 \tan k_1d)  \cos^2 k_1d
             \ . 
\end{equation}
From (43) and (44), we write
\begin{equation}
 B-A = {\mu_1 \over k_2}(k_1 \tan k_2c  + k_2 \tan k_1d)  \ .
\end{equation}
Multiplying side by side relations (51) and (52) and using the identity
\begin{equation}
 \Gamma^2 (B-A)^2 = (1+ \Gamma^2 B^2) +(1+ \Gamma^2 A^2)-
                  2(1+ \Gamma^2 AB) \ ,
 \end{equation}
we find
\begin{equation}
 2(1+ \Gamma^2 AB) = (1+ \Gamma^2 A^2) 
             \left[
                 1+{ 1+ \Gamma^2 B^2 \over 1+ \Gamma^2 A^2}
                 -{W \cos^2k_1d \over k_1k_2}
             \right] \ ,
\end{equation}
where
\begin{equation}
 W= (k_1 \tan k_1d + k_2 \tan k_2c) (k_1 \tan k_2c + k_2 \tan k_1d) \ .
\end{equation}
Using (48), (54) turns out to be
\begin{equation}
 {1+ \Gamma^2 AB \over 1+ \Gamma^2 A^2}=
              {1 \over 2} 
             \left[
                 1+{ \cos^2 k_1d \over \cos^2 k_2c}
                 -{W \cos^2k_1d \over k_1k_2}
             \right] \ ,
\end{equation}
Substituting this result in (50), we find
\begin{equation}
\cos Ke={1 \over 2} 
             \left[
                 1+{ \cos^2 k_1d \over \cos^2 k_2c}
                 -{W \cos^2k_1d \over k_1k_2}
             \right]
          {{\cos k_{2}c}\over{\cos k_1d}}
\end{equation}
Using expression (55) of $W$, this last relation leads 
straightforwardly to (23).

Let us now consider the case where $E<V_{0}$ and set 
\begin{equation}
k_{3}={\sqrt{2m(V_{0}-E)} \over \hbar}  \ .
\end{equation}
In Refs. \cite{NGO,Sin}, by using the continuity of the wave function 
and its derivative, it is shown that 
\begin{equation}
\cos Ke =
       \cosh\left(k_{3}d\right)\cos\left(k_{2}c\right)-
        {{k_{2}^2-k_{3}^2}\over{2k_{2}k_{3}}}
        \sinh\left(k_{3}d\right)\sin\left(k_{2}c\right)
\end{equation}
This relation was obtained for the first time by Kr\"onig and Penney 
\cite{KP}. As it is the case for (23), an investigation of (59) shows 
the existence of a band structure for the energy 
spectrum. In order to reproduce it with the use of the the reduced 
action properties, let us choose as independent 
real solutions of the Schr\"odinger equation, Eq. (4),  
the two couples 
\begin{equation}
(\sinh k_3x, \cosh k_3x), \hskip20mm (\sin k_2x, \cos k_2x)
\end{equation}
respectively in regions $I$ and $II$. With the use of (30), we have
\begin {equation}
S_0^I(x)=\hbar \ \arctan {\left[ \mu_3 \tanh (k_3 x)  +\nu_3  \right]} 
           +\hbar l_3 \; ,
\end {equation}
and $S_0^{II}(x)$ keeps the same expression as in (33). By 
appealing to the continuity conditions (24), (25) and (26) at 
$x=0$ for the reduced action and its derivatives, and to Bloch 
periodicity condition with its derivatives, Eqs. (27), (28) and 
(29), and by following the same procedure as above, we 
get to relation (59) without using the wave function.

We would like to emphasize that, in order to reproduce the 
dispersion relations (23) and (59), the choice of the couples 
(31) and (60) of solutions of the Schr\"odinger equation used in 
the reduced action is not an essential. Since the Schr\"odinger 
equation is linear, other choices which must be linear combinations 
of the above solutions are also possible. However, any other choice 
must reproduce the same dispersion relations. In fact, as shown in 
Ref. \cite{BD2}, we can check that the reduced action is 
invariant under any linear transformation of the solutions of 
the Schr\"odinger equation by redefining suitably the integration 
constants $(\mu, \nu, l)$.

\section{Conclusion}

The present work can be summarized in two main results.
\begin{enumerate}
 
\item In a periodic potential case, we established the 
condition of Bloch periodicity for the reduced action 
by using the relation between the wave function and the reduced 
action established in the context of the equivalence postulate of 
quantum mechanics. The analogous version of this theorem in Bohm's 
theory is also deduced.
 
\item  In this context, by using the quantum Hamilton-Jacobi equation, 
we also reproduced the well-known dispersion relations which 
predict a band structure for the energy spectrum
without appealing to the wave function or to its
usual axiomatic interpretation. These relations can be also reproduced  
in the context of the Bohm theory which appears here as a particular 
case in which we take $(\alpha =1, \beta=0)$ and then 
$(\Gamma=1, \Delta=0)$.

\end{enumerate}

To conclude, we think that the present work is a further argument 
to reinforce the belief that the equivalence postulate of 
quantum mechanics constitutes a serious alternative to the standard 
quantum mechanics. In fact, firstly it allows to reproduce the 
well-known results 
as it was already the case both for the tunnel effect and energy 
quantization \cite{FM2,FM3}.    
Secondly, it provides an appropriate frame to reconcile 
general relativity with quantum mechanic.

\vskip\baselineskip
\noindent
{\bf REFERENCES} 

\end{document}